\documentclass[conference]{IEEEtran} 

\usepackage{algorithm}
\usepackage[noend]{algorithmic}
\usepackage[geometry]{ifsym}
\usepackage{ifsym}
\usepackage{amssymb}
\usepackage{amscd}
\usepackage{dsfont}
\usepackage{amsmath}
\usepackage{epsfig}
\usepackage{graphics}
\usepackage{psfrag}
\usepackage{rotating}
\usepackage{amsmath}
\usepackage{amsfonts}
\usepackage{url}
\usepackage{epstopdf}
\usepackage{color}
\usepackage[usenames,dvipsnames]{xcolor}
\usepackage{mdwlist}
\usepackage[printonlyused]{acronym}

\usepackage{hyperref}
\usepackage{subfigure}
\IEEEoverridecommandlockouts	%enable \thanks

\usepackage[footnote,nomargin]{fixme}

\newcommand{\fnfm}[1]{\fxnote{FM: #1}}
\newcommand{\fnsv}[1]{\fxnote{SV: #1}}

\graphicspath{{src/}}
\newcommand{\fref}[1]{Fig.~\ref{#1}}
\newcommand{\sref}[1]{Sec.~\ref{#1}}

\newcommand{\eref}[1]{(\ref{#1})}

\newcommand{\scalefactorplot}{0.7}
\newcommand{\scalefactorplottwo}{0.8}

\hypersetup{draft=true} %needed according to IEEE regulations

\begin{document}

\title{Predicting a User's Next Cell With Supervised Learning Based on Channel States}

\author{Xu Chen\IEEEauthorrefmark{1}\IEEEauthorrefmark{3}\thanks{This work was supported by the program \emph{RISE professional} of the DAAD and by the \emph{PhD@Bell Labs} internship program.}, Fran\c cois M\'eriaux\IEEEauthorrefmark{2}\IEEEauthorrefmark{3}, and Stefan Valentin\IEEEauthorrefmark{3}\medskip\\
\IEEEauthorrefmark{1}Department of Electrical Engineering, Princeton University, NJ, USA\\
\IEEEauthorrefmark{2}Laboratoire des Signaux et Syst\`emes, SUPELEC, Paris, France\\
\IEEEauthorrefmark{3}Bell Labs, Alcatel-Lucent, Stuttgart, Germany\\
{\small xuchen@princeton.edu, meriaux@lss.supelec.fr, and stefan.valentin@alcatel-lucent.com}
}

% use \IEEEauthorrefmark{n} to distinguish different affiliations
\maketitle
%\vspace{-0.55cm}
%\pagestyle{empty}
%\thispagestyle{empty}

\begin{abstract}
Knowing a user's next cell allows more efficient resource allocation and enables new location-aware services. To anticipate the cell a user will hand-over to, we introduce a new machine learning based prediction system. Therein, we formulate the prediction as a classification problem based on information that is readily available in cellular networks. Using only Channel State Information (CSI) and handover history, we perform classification by embedding Support Vector Machines (SVMs) into an efficient pre-processing structure. Simulation results from a Manhattan Grid scenario and from a realistic radio map of downtown Frankfurt show that our system provides timely prediction at high accuracy.
\end{abstract}

\maketitle

\section{Introduction}
Localizing a user and tracking its trajectory change today's cellular networks. At application layer, many services make heavy use of the user's current position. At the physical layer, accurate tracking is crucial for many beamforming approaches.

In this paper, we focus on a coarse localization of the wireless user. Instead of obtaining accurate geographical coordinates, we are satisfied in expressing the user's location in terms of a cell index. However, rather than obtaining the user's current location or a short-term trajectory, we propose a framework to \emph{predict} the next cell the user will hand-over to. Although this prediction is performed at low geographical accuracy and at large time scale, it has to be provided early enough to allow the system to adapt. This requires an accurate prediction before the user enters the next cell, thus, rendering simple, linear predictors inapplicable with many applications.

Such applications include various adaptation functions at the medium access and network layer. New, \emph{context-aware} schedulers can provide seamless quality-of-service by knowing in advance that a user will join a congested cell \cite{proebster11:cara_journal}. Interference management schemes can blank subframes of interfering users even before they join an interfered cell and context-aware handover schemes can use our long-term prediction to make faster and better decisions.

Unlike previous work, based on the Global Positioning System (GPS) or Time Difference Of Arrival (TDOA) \cite{Elnahrawy2004}, our solution combines Channel State Information (CSI) with limited handover history. This combination of long-term handover information with short-term CSI, has several benefits:
\begin{itemize*}
\item It is fully functional within buildings. This is not the case for GPS-based solutions.
\item It employs information that is readily available in cellular networks. As handhelds perform frequent CSI feedback and the network stores handover history, no signaling costs are added to the radio link and implementation effort is limited.
\item It predicts the user's next cell early and at high accuracy. This high performance was observed in a synthetic and in a realistic scenario, without strong assumptions on user mobility.
\end{itemize*}

Our paper is structured as follows. We discuss related approaches in \sref{sec:relwork} and introduce the system model in \sref{sec:sysmod}. \sref{sec:framwork} details the proposed prediction framework, which is studied in \sref{sec: simu results}. \sref{sec:concl} concludes the paper.

\section{Related Work}
\label{sec:relwork}
Two fields in literature are related to our work: predicting the user's trajectory and estimating its current location. Some work on localization \cite{Elnahrawy2004,Kim2009,Kejiong2012} uses CSI as input or shares the application of Support Vector Machines (SVMs) \cite{Tran2008} with our approach. In contrast to this work, we complement CSI by the users' handover history to predict the user's next cell. Thus, our approach differs in objective as well as in input from the state-of-the-art in localization.

Following our objective, \cite{Kyriakakos2003,Anagnostopoulos2007} employ machine learning to predict a user's next cell. These predictors are solely based on handover history, while our work combines this information with CSI. Both types of information are efficiently available in cellular networks but substantially differ in time scale. While CSI is periodically reported within several milliseconds \cite{3gpp10:lte_mac_spec}, obtaining a sufficiently long vector of previous handovers requires minutes or, for slow users, hours. 

Unlike handover history, large CSI vectors can be obtained quickly and capture the scatterers in a user's propagation environment. We will see that such CSI ``fingerprints'' provide substantially higher prediction accuracy than handover history alone.

\section{System Model}
\label{sec:sysmod}
We study a cellular network with $\mathcal{I}=\{1,\ldots,I\}$ users and a base station (BS), under the following modeling assumptions.

\subsection{Channel Model}\label{sec: channel model}
We assume that the wireless communication channel of user $i \in \mathcal{I}$ within a discrete time slot $t$ is characterized by the channel gain 
\begin{equation}
\label{eq:csi}
g_i(t) = |h_i(t)|^2 d_i(t)^{-\alpha(x_i(t))}.
\end{equation}
To account for long-term propagation loss, $d_i(t)$ denotes the distance between user $i$ and BS. The path-loss exponent is denoted by $\alpha(x_i(t))$ and depends on the user's geographical position $x_i(t)$.

Fast fading is reflected by $|h_i(t)|^2$, which we assume to be an exponentially distributed random variable that is independent among the users. We adopt the Jakes-like fading model where the time-correlation of $d_i(t)$ follows the zeroth-order Bessel function of the first kind, which is parametrized by the maximum Doppler shift defined by the user's speed. This model is widely used to reflect an isotropic scattering environments or ensembles of multiple environments \cite[Sec. 2.4.3]{tse05:wireless_comm}.

\subsection{CSI Feedback}
High-rate communication systems such as IEEE 802.16 and LTE, adapt the modulation scheme, code rate, and allocated time-frequency blocks to the current state of the wireless channel. As in those systems the channel cannot be assumed to be reciprocal, users perform periodic feedback of CSI. To accurately adapt to a time-variant channel, this reporting should be performed once per channel coherence time, i.e., between 1 and 20\,ms in many scenarios. While typically a quantized channel state is reported, we assume the reporting of the channel gain $g_i(t)$ for tractability.

\subsection{Mobility Model}
\label{sec:mobmod}
We assume that a user's mobility is fully described by its speed and its motion path. In each cell, we assume that there is a finite amount of paths each user can take. Each path has an entry point and an exit point in the cell, where the path is a continuous line joining those points. Paths may have common properties, such as the same entry point, same exit point, or even complete sections. 

To select a user's path and speed, we adopt a random model. Any time a users enters a cell it randomly chooses a speed value from a given interval and a path value from a given set (cp. \sref{sec: simu results}). The choices are independent among the users and follow a uniform distribution. Consequently, path and speed may change at any handover.

While moving on a given path in the cell, a user $i \in \mathcal{I}$ reports the channel gain $g_i(t)$ to the base station. According to \eref{eq:csi}, $g_i(t)$ depends on $|h_i(t)|^2$, representing fast fading, and on $d_i(t)^{-\alpha(x_i(t))}$, which accounts for slow fading. As this slow fading term depends only on the user's location, it is fully determined by the user's path. However, even if multiple users travel on the same paths, it is very likely that their CSI sequences differ due to different speeds and fast fading.\fnsv{the problem with this channel model is that fast fading has no effect (averages out) and that scatterers are ignored.}

In Sec.~\ref{sec: simu results}, two maps are used to perform simulations to test the proposed trajectory prediction. The first map is a Manhattan grid. It is an abstract map in which the path-loss exponent $\alpha$ is set arbitrarily to model different shadowing effects in different areas of the map. In the second map, however, this function is not set arbitrarily since $d_i(t)^{-\alpha(x_i(t))}$ is obtained from real measurements in the German city Frankfurt am Main.

\section{Proposed Prediction Framework}
\label{sec:framwork}
We propose to predict the user's next cell by solving a classification problem via supervised learning. To do so, we build classifiers from the user's CSI sequence as it travels through the current cell and from the index of its next cell. While this index is used as a label, the CSI sequence serves as an input vector. Before we describe the real-time prediction during system operation, let us formulate the classification problem.

\subsection{Next Cell Prediction as Classification Problem}
\label{sec:Classification_Problem}
A user's trajectory can be represented by a set of cells it travels through. Focusing on the cell an arbitrary user $i$ is currently associated with, we define a user's trajectory using the index $p_i$ of its previous cell and index $n_i$ of its next cell. Furthermore, the trajectory can be associated to CSI which is reported periodically while the user is traversing its current cell. Formally, this CSI sequence can be stated as
\begin{equation}
H_i \triangleq  \left\{ \begin{array}{ll}
         \emptyset & \mbox{if $t < t_i^ \text{in}$ };\\
        (g_i(t^\text{in}_i), \ldots , g_i(t)) & \mbox{if $t^\text{in}_i \leq t < t^\text{out}_i$};\\
         (g_i(t^\text{in}_i), \ldots , g_i(t^\text{out}_i)) & \mbox{if $t \geq t^\text{out}_i$}.\\
         \end{array}\right.
\end{equation}
Here, CSI is represented as the channel gain $g_i(t)$, $t^\text{in}_i$ denotes the time when user $i$ enters the current cell, and $t^\text{out}_i$ is the time when the users leaves this cell.

Given the input data $H_i$ and $p_i$, we can predict the user's next cell by finding an accurate mapping between the tuple $\langle p_i,H_i\rangle$ to the associated next cell $n_i$. As there will only be a limited number of possible combinations of adjacent cell indices, this is a classification problem and can be solved by supervised learning methods. 

The benefit of using this input information is that it is readily available in current cellular networks. The CSI vector $H_i$ is known at the user and reported to the base station. The previous cell index is known at user and base station after the handover. This simplifies the implementation of our framework either at the base station or at the handset.

In our approach, multiple classifiers are learned. Each classifier corresponds to a possible previous cell and each predicts a next cell out of all possible next cells. Consequently, the number of classifiers is equal to the number of possible previous cells. Compared to using only a single classifier, a learning process using multiple classifiers is faster and the prediction is more accurate. This is because each single classifier only needs to perform a much smaller-scale classification task.

\begin{figure}
	\begin{center}
	\includegraphics[width=\columnwidth]{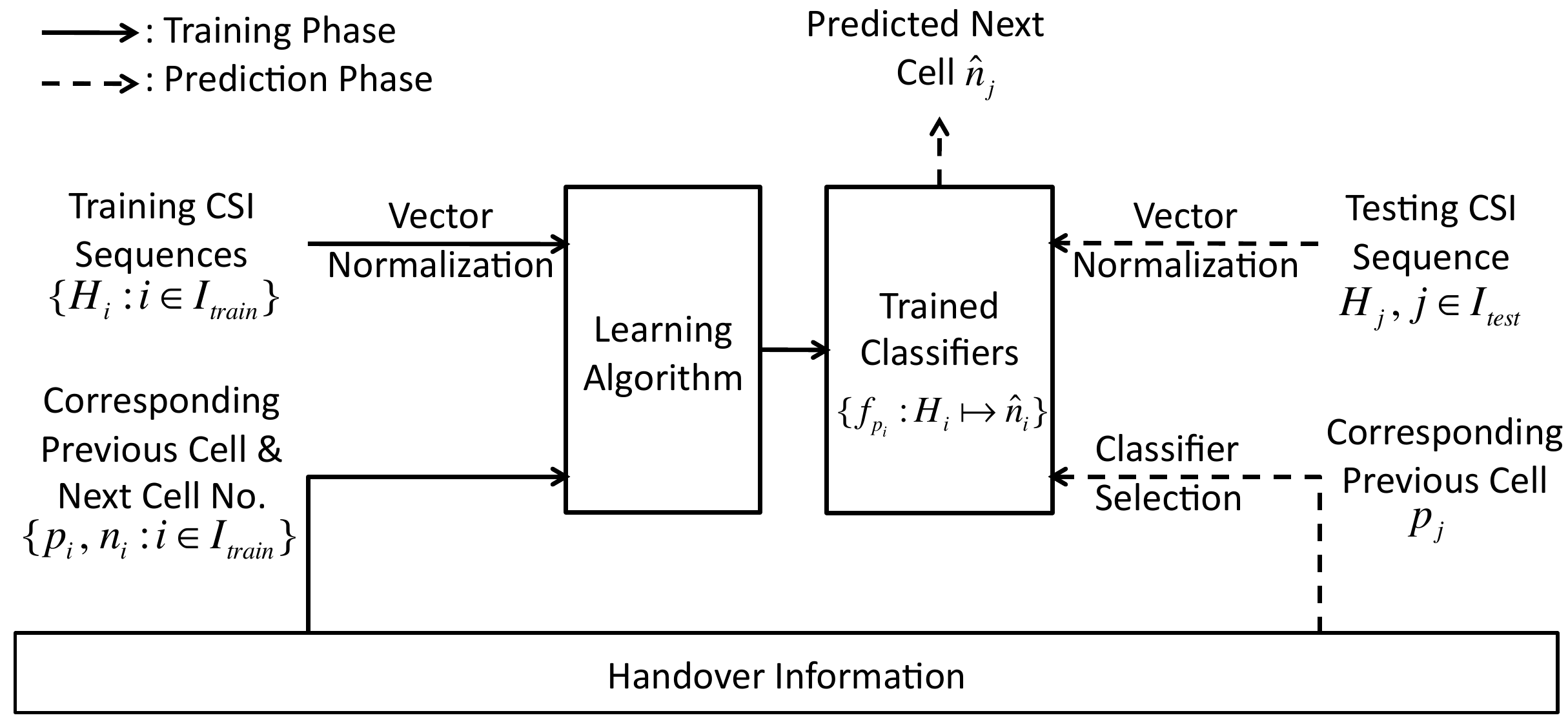}
	\end{center}
\caption{Concept of the proposed prediction scheme}
\label{fig:Offline_HO_Scheme} 
\end{figure}

This idea combined with supervised learning is illustrated in Fig.~\ref{fig:Offline_HO_Scheme}. The overall scheme consists of two phases, training phase and prediction phase. In the training phase, the first input to the system is a set of training CSI sequences $\{H_i: i \in \mathcal{I}_\text{train}\}$, their associated previous cell $\{p_i\}$ and their next cell indices $\{n_i\}$. Here, $p_i$ specify particular classifiers and $\{n_i\}$ are the corresponding labels that the learning algorithm aims to match the CSI sequences $\{H_i\}$ to. Note that a vector normalization function applies subsampling and interpolation to assure that the CSI sequences are of equal length.

Based on this training data, a learning algorithm derives multiple classifiers $\{f_{p_i}\}$ corresponding to all the possible previous cells. Using this result of the training phase, the prediction phase now selects a specific classifier $f_{p_j}$ to match a previously unknown CSI sequence $H_j, \forall j \in \mathcal{I}_\text{test}$, to a predicted index of the next cell $\hat{n}_j$. 

%{\color{Plum}Before either training or testing CSI sequences are fed to the scheme, preprocessing techniques such as interpolation and subsampling are applied, so as to normalize all input vectors to the same length.}

\subsection{Classification}
\label{sec:SVM}
We employ a multi-class Support Vector Machine (SVM) as a classifier, using the LIBSVM \cite{Chang2001} library. 

SVMs \cite{Vapnik1995} seek for the decision boundary between any two classes by constructing a hyperplane in a high-dimensional space such that the hyperplane has the largest distance to the nearest training sample of either class. Although a linear SVM is fast to train, in our case, it does not achieve an accurate early prediction, when feature dimension is small. This situation is common when users traverse the cells faster (e.g., at vehicular speed). To this end, we choose a nonlinear SVM with a Gaussian kernel $K(\textbf{x}, \textbf{y}) = \exp{(-\gamma {\|\textbf{x}-\textbf{y}\|}^2)}$ that maps
sample vectors $\textbf{x}, \textbf{y}$ into a higher dimensional space. The Gaussian kernel parameter $\gamma$ and the SVM slack variable $C$ are optimized with cross-validation on a subset of the training set based on grid search.

\subsection{Real-Time Prediction Scheme}
\label{sec:Online_Scheme}
Based on the above classification problem, we now describe a system that predicts the user's next cell in real time during system operation. The main idea is to predict a user's next cell from current classifiers that are learned from all user's traversing the cell. After a user has left the cell, the true next cell index is known and the classifier can be updated. This feedback approach includes more and more users in the training set and increases prediction accuracy over time. By frequently renewing the training set and re-training the classifiers, the prediction is continuously updated during system operation.

\begin{figure}
	\begin{center}
	\includegraphics[scale=0.32]{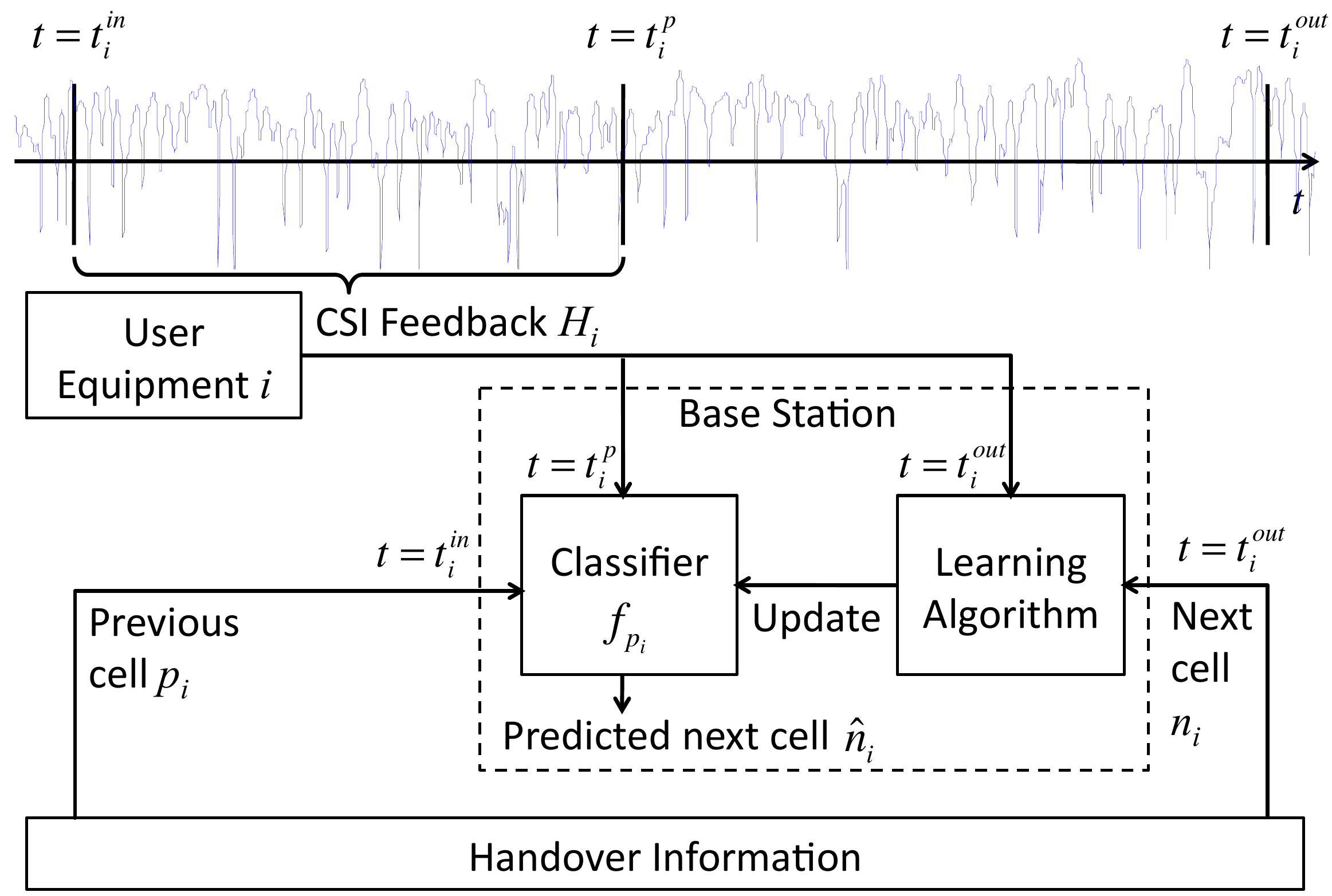}
	\end{center}
\caption{Real-time system integration of the proposed prediction scheme}
\label{fig:Online_Scheme} 
\end{figure}

We illustrate this online prediction scheme in \fref{fig:Online_Scheme} for an arbitrary user $i$. The user enters the current cell at entering time $t^\text{in}_i$ and exits at time $t^\text{out}_i$. Additionally, we define the time when the prediction is performed at $t^{p}_i$.

This leads to two time intervals. Interval $[t^\text{in}_i, t^{p}_i]$ defines the duration over which the input data is obtained until the prediction is performed. Interval $[t^{p}_i, t^\text{out}_i]$ represents the time the predictor can look ahead. 

When the prediction is performed at $t^{p}_i$, the CSI sequence $H_i$ observed during $[t^\text{in}_i, t^{p}_i]$ is used with classifier $f_{p_i}$ to derive the index of the next cell $\hat{n}_i$. After user $i$ exits the current cell at time $t^\text{out}_i$, the true next cell index $n_i$ as well as CSI sequence $H_i$ will be fed back to the learning algorithm and thus update the corresponding classifier $f_{p_i}$.

The smaller $t^{p}_i$, the less input data is required and the earlier the prediction can be used by network adaptation algorithms. Performing the adaptation at $t^{p}_i+\epsilon$ -- allowing a very small $\epsilon$ for computation time -- assures that prediction and adaptation are performed in real time. We will study these timing aspects in the following section.

\section{Simulation Results}
\label{sec: simu results}
To obtain tractable results, we first study our system in a simple Manhattan grid scenario. Then, we move to a realistic radio map for the German city Frankfurt am Main. For both scenarios, our results consistently show high performance and fast convergence for our prediction framework.

\subsection{Manhattan Grid Scenario}
\begin{figure}
\begin{center}
\includegraphics[width=\columnwidth]{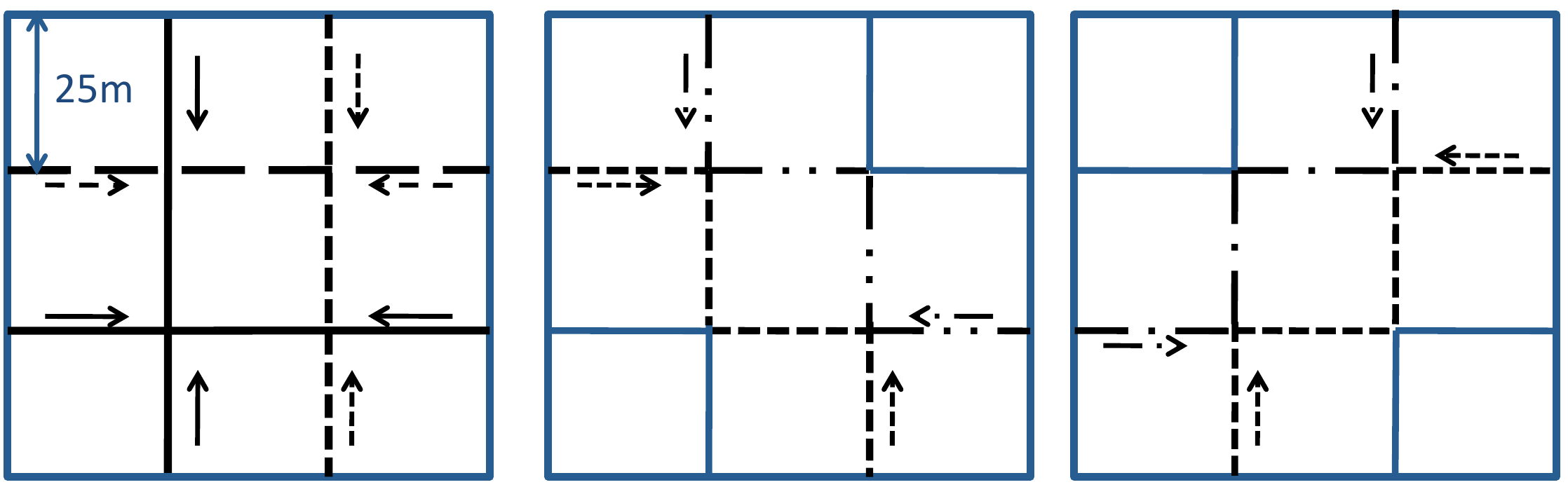}
\end{center}
\caption{Illustration of the Manhattan Grid scenario: A quadratic area of $5625\,\textnormal{m}^2$ size}
\label{fig:Manhattan_Map}
\end{figure}

We design this simple scenario based on quadratic cells, as illustrated in Fig.~\ref{fig:Manhattan_Map}. Each cell is of $75\,\textnormal{m} \times 75\,\textnormal{m}$ size and includes 16 distinct paths. As all cells are assumed to be equal, we can focus on a single cell without loss of generality. This exemplary cell is surrounded by 4 neighboring cells at each cardinal direction. Hence we have four different labels for the users moving through this cell. 

The paths within the studied cell traverse the cell either horizontally, vertically, or diagonally. The users choose the path randomly when they enter the cell and follow the path at a speed, randomly chosen from between 5 m/s and 40 m/s. This random choice follows a uniform distribution and all users make independent choices, as described in \sref{sec:mobmod}.

To evaluate the accuracy of the proposed solution, we first generate classifiers in the learning phase. To assure accurate learning of the classifiers, we collect 500 data samples per path and use $90\%$ for training and $10\%$ for testing. 
%Multiple rounds of cross-validations are performed to ensure the generality of the estimated accuracy.

\begin{figure}
\begin{center}
\subfigure[Prediction accuracy versus sample length ratio]{
\includegraphics[width=\scalefactorplot\columnwidth]{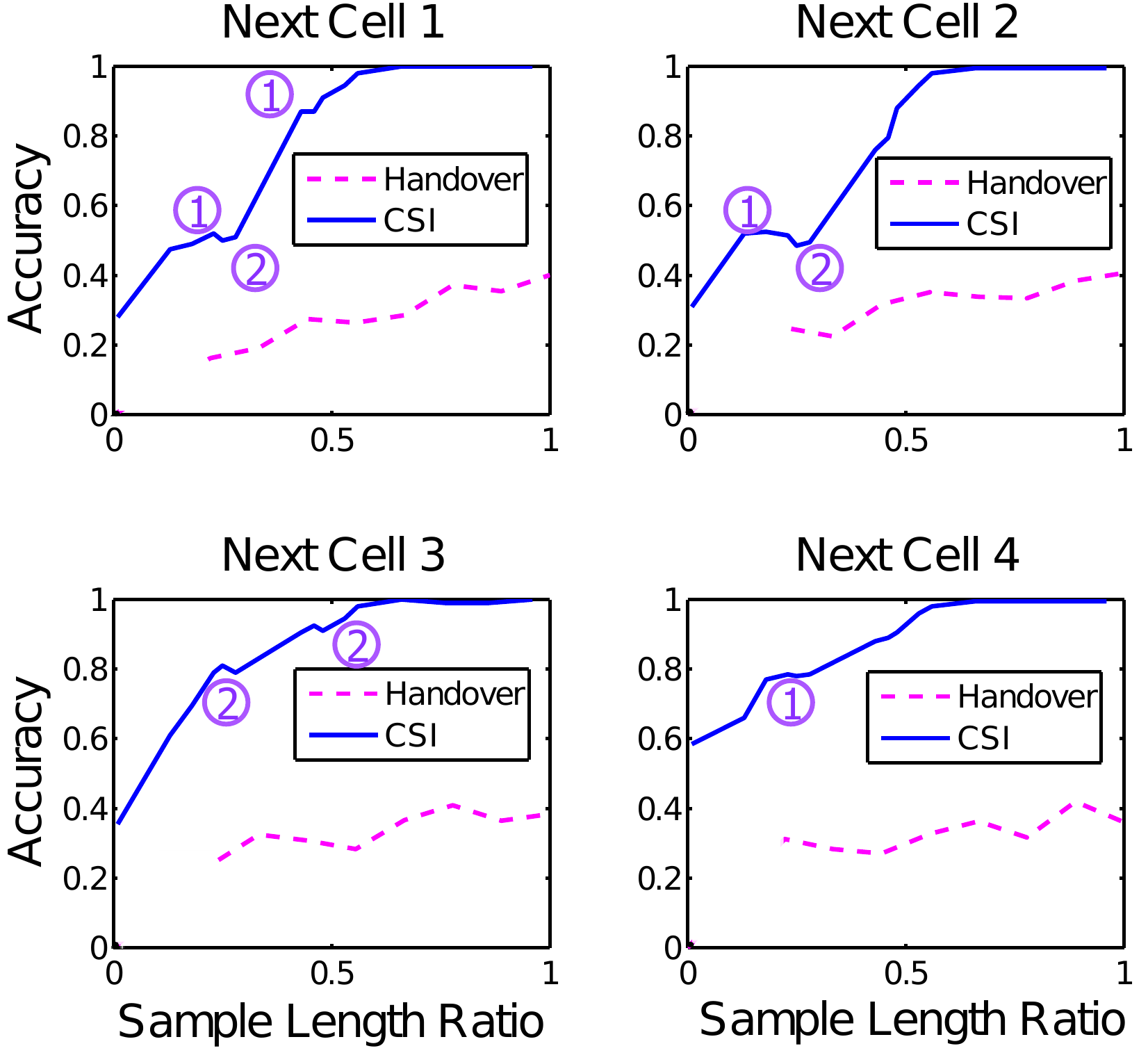}
\label{fig:Ac_SLR_Manhattan} } 
\subfigure[Prediction accuracy versus simulated time]{
\includegraphics[width=\scalefactorplot\columnwidth]{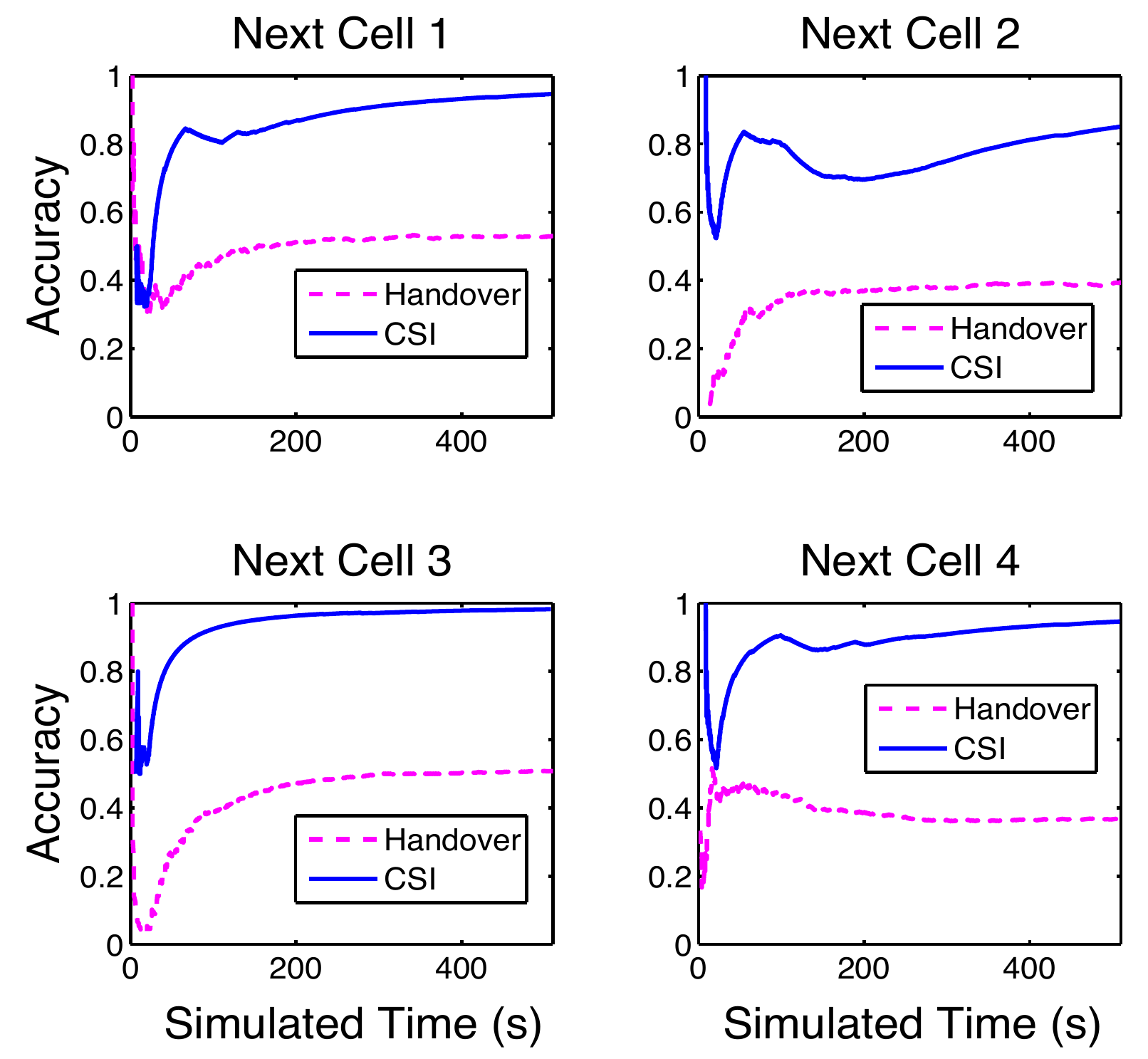}
\label{fig:Ac_Online_Manhattan} } 
\end{center}
\caption{Prediction accuracy of the proposed scheme and the baseline in the Manhattan Grid scenario}
\label{fig:Manhattan_Result}
\end{figure}

Fig.~\ref{fig:Manhattan_Result}\subref{fig:Ac_SLR_Manhattan} presents the prediction accuracy for each label, i.e., the next cell the user is predicted to go to. The results are shown for a varying sample length ratio, which represents the fraction of input data used from the available data. For our scheme, this is the proportion of CSI values used out of all CSI values that a user reports in the current cell. 

We form a baseline by applying SVM only on the user's handover history. Here, the sample length ratio is the length of the used handover history divided by the absolute length of the history. We generate handover history using the random mobility model from \sref{sec:mobmod}. With this model, all possible next cells have the same probability of being part of a trajectory, if the history length is sufficiently large. We, thus, consider no history length smaller than 2, i.e., a sample length ratio below $0.25$. \fnsv{wrote it more direct. hope it's still correct.}

As shown in \fref{fig:Manhattan_Result}\subref{fig:Ac_SLR_Manhattan}, 
%{\color{Plum}the prediction accuracy increases with the sample length ratio of CSI sequence, while as expected, the accuracy floats around 25\% when the used handover history length changes.\fnsv{wrong. please look at the figure: it goes up to 40\%} 
our method substantially outperforms the handover history-based approach. This illustrates that older handover values do not accurately reflect the decision of the user's current trajectory. Also, our method is shown to profit quickly from more input CSI data, it only requires a small fraction of input data to reach a certain accuracy threshold.

In Fig.~\ref{fig:Manhattan_Result}\subref{fig:Ac_SLR_Manhattan}, we notice several intervals with slowly increasing accuracy. We mark those phases by (1) and use (2) to mark phases where the accuracy even decreases for higher sample ratios. Studying the Euclidean distance among the paths explains both effects as a result of the overlapping sections of different paths. As different paths can use the same section of a street, the long-term average of the CSI values for this section is similar. Unless the paths separate, using a large sample length ratio does not increase prediction accuracy. 

Nonetheless, the prediction accuracy reaches 1 when the sample length ration exceeds $0.6$. This means that once a user travels $60\%$ of its path through the cell, maximum prediction accuracy is reached. With the considered speeds, this allows to predict the next cell at 100\% accuracy between 1\,s and 20\,s in advance. This provides sufficient time for many adaptation schemes before the user leaves the cell. 

\fref{fig:Manhattan_Result}\subref{fig:Ac_Online_Manhattan} shows prediction accuracy per label for varying simulated time. These results where generated for the real-time  prediction scheme with users entering the current cell with Poisson distributed inter-arrival times at $\lambda = 1$\,s. We notice that prediction accuracy converges very fast. With approximately 100 samples per path, our learning scheme obtains stable classifiers. This fast convergence to 100\% accuracy shows, again, the advantage of our proposed CSI-based prediction over using handover history alone.

\subsection{Frankfurt Scenario}
\begin{figure}
\begin{center}
\subfigure[Base station locations]{
\includegraphics[width=0.44\columnwidth]{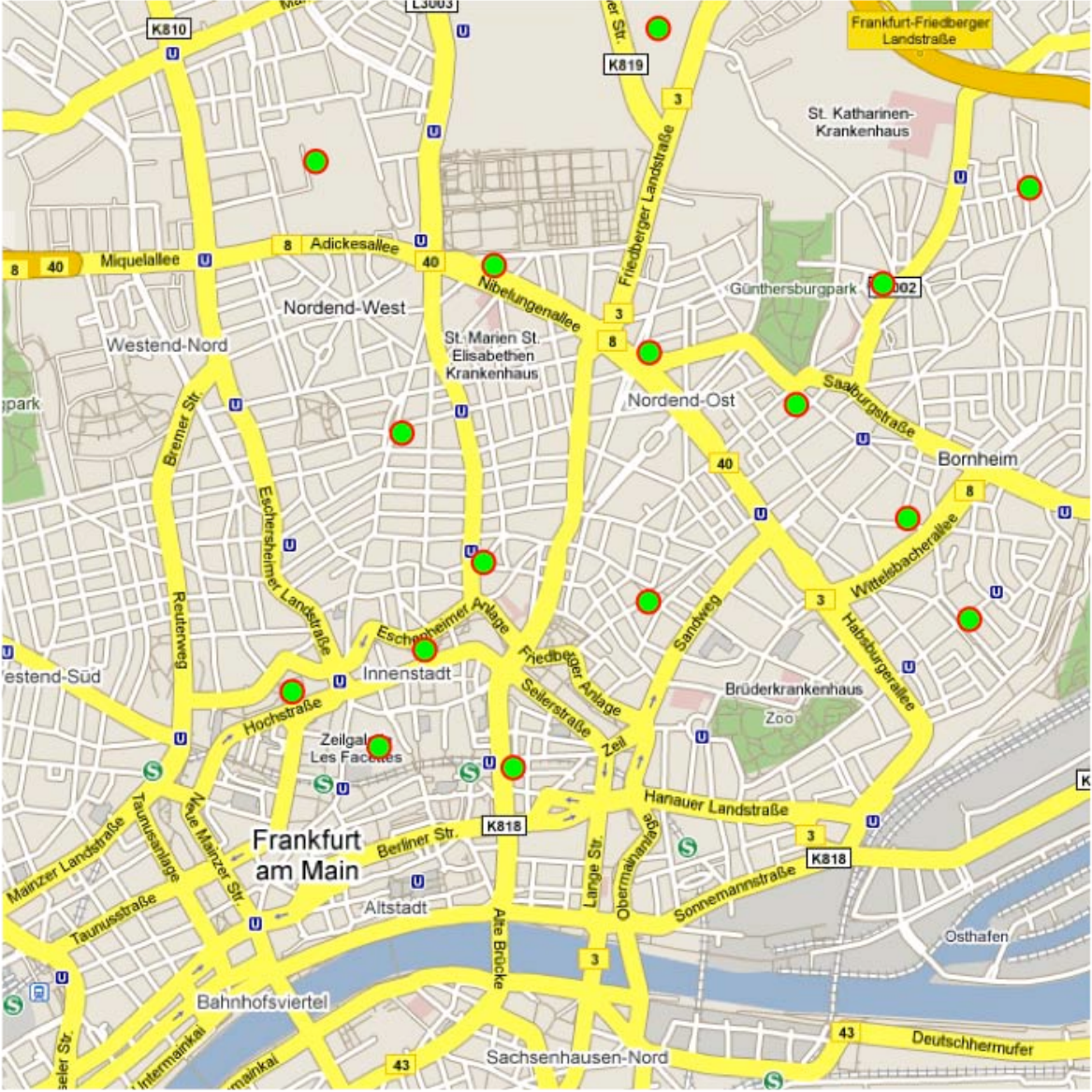}
\label{fig:CityMap} } 
\subfigure[Cell distribution and studied cells]{
\includegraphics[width=0.49\columnwidth]{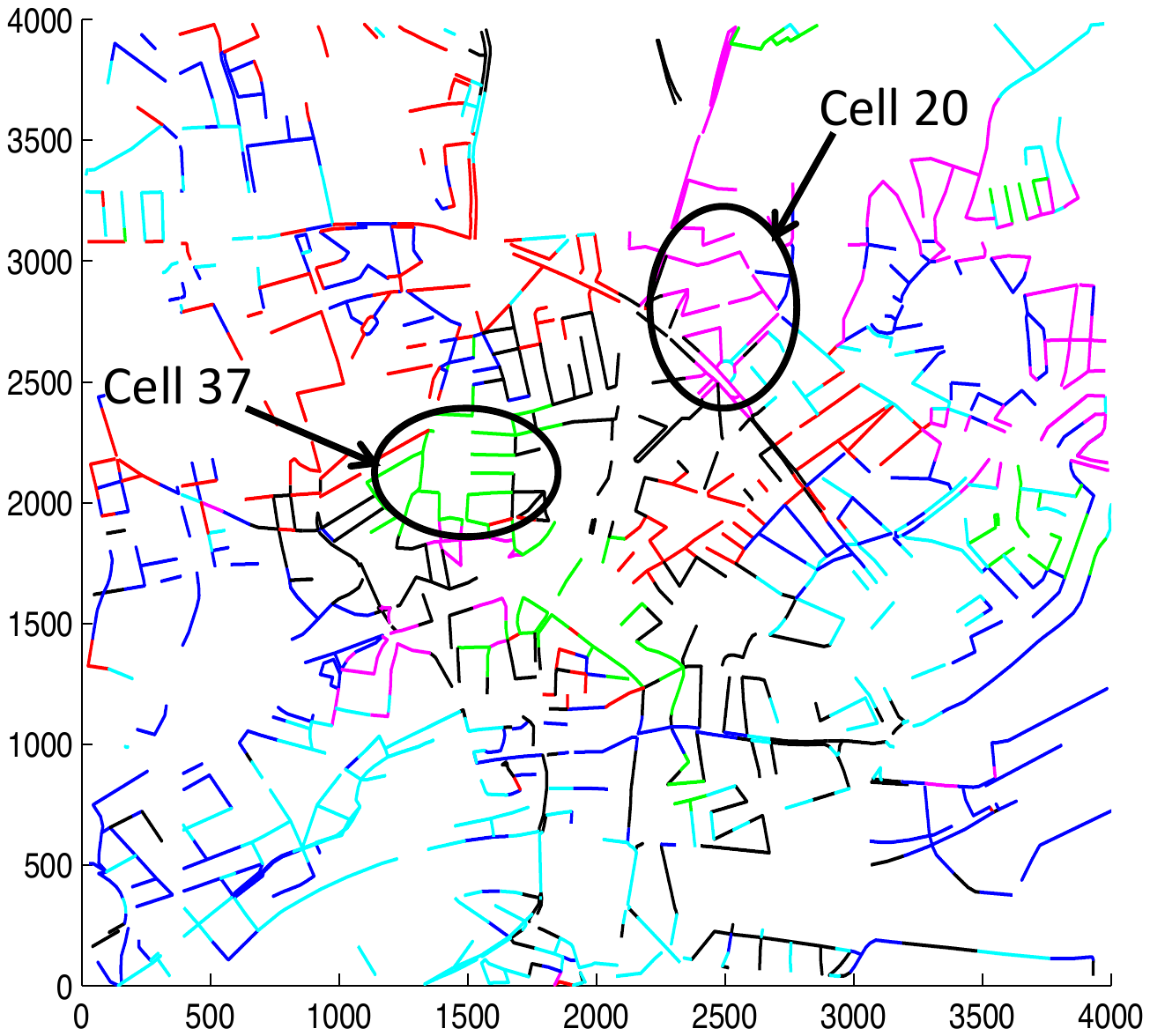}
\label{fig:ColorPath} } 
\end{center}
\caption{Overview of the Frankfurt scenario: A $16\,\textnormal{km}^2$ area in downtown Frankfurt am Main, Germany}
\label{fig:Frankfurt_Map}
\end{figure}

In this scenario, we study an area of $4\,\textnormal{km} \times 4\,\textnormal{km}$ in the downtown area of the German city Frankfurt am Main. A map and the 16 base station locations are shown in \fref{fig:Frankfurt_Map}\subref{fig:CityMap}. As each base station has 3 sectors, drive tests in this area recorded the average channel gain from 48 cells. Using those measurements to parametrize our fast fading model, \sref{sec: channel model}\fnfm{moved the comma} provides the channel gain $H_i$ for user $i \in \mathcal{I}$.

This scenario differs from Manhattan grid by its more complicated cell geometry and trajectories. However, the remaining parts of the system model are not changed. The baseline could not have\fnfm{have was missing} been studied in this scenario due to insufficient handover data.

\begin{figure}
\begin{center}
\subfigure[Results for Cell 37]{
\includegraphics[width=\scalefactorplot\columnwidth]{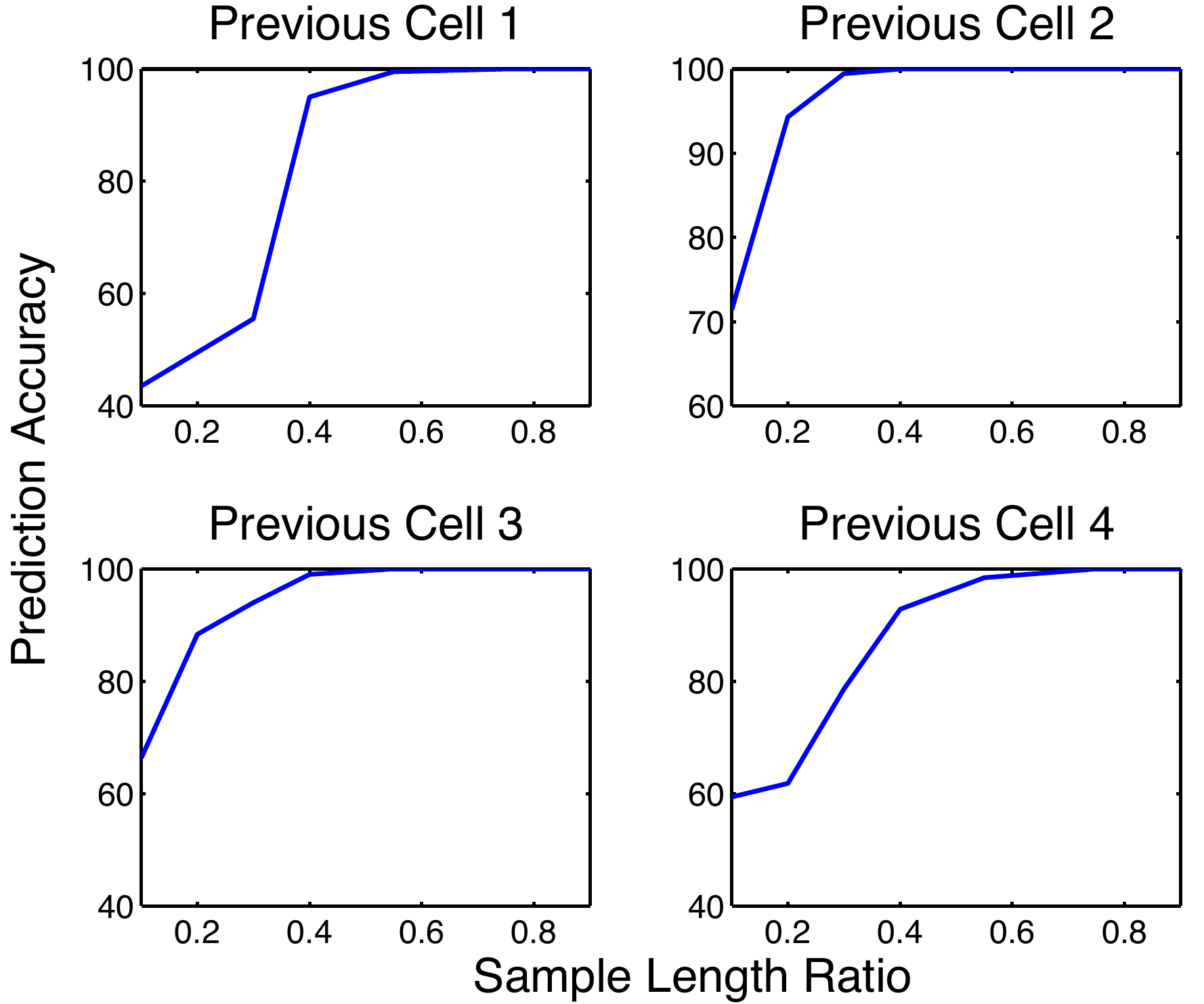}
\label{fig:Cell37} } 
\subfigure[Results for Cell 20]{
\includegraphics[width=\scalefactorplottwo\columnwidth]{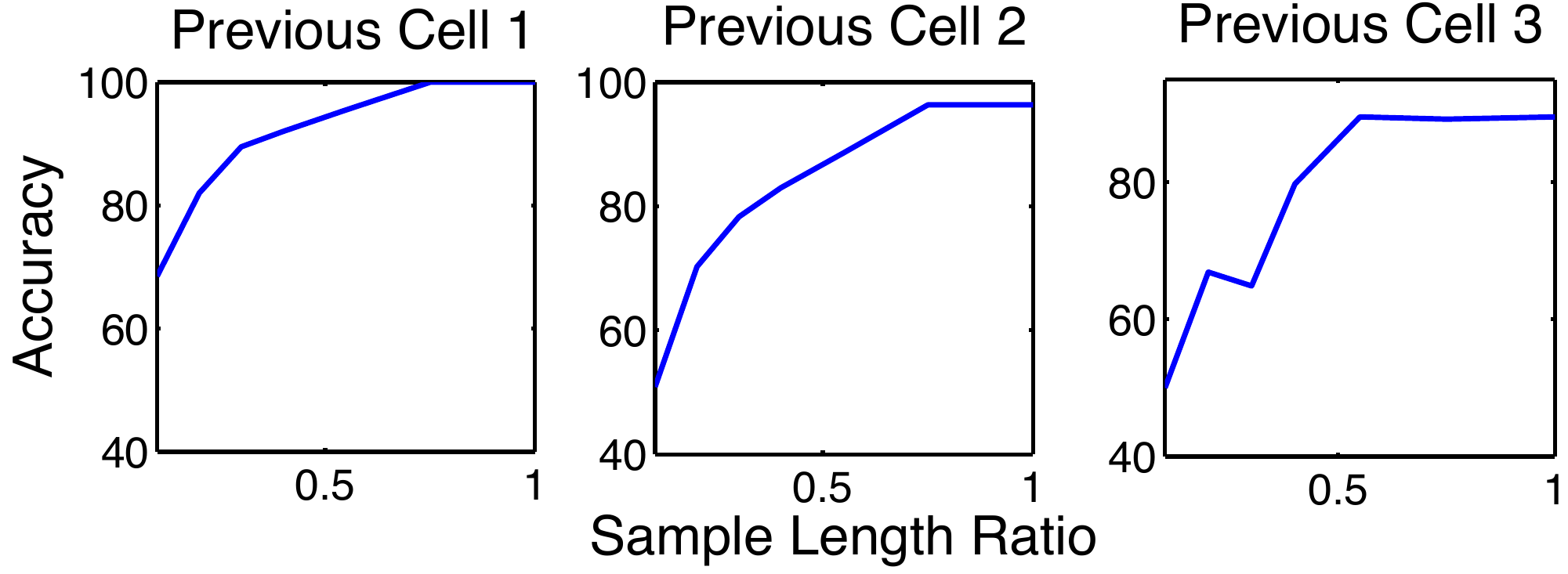}
\label{fig:Cell20} } 
\end{center}
\caption{Prediction accuracy of the proposed scheme in the Frankfurt scenario}
\label{fig:Frankfurt_Result}
\end{figure}

We test our predictor for two different cells, which are marked in \fref{fig:Frankfurt_Map}\subref{fig:ColorPath}. The first cell with index 37 includes 26 paths. Here, 100\% prediction accuracy is reached when more than 70\% of the CSI input values are used; cp. \fref{fig:Frankfurt_Result}\subref{fig:Cell37}. With the second cell, number 20, we study a complicated example. This cell contains 46 paths and has a more complex geometry than Cell 37. As shown in \fref{fig:Frankfurt_Result}\subref{fig:Cell20}, the prediction accuracy does not reach $100$\% even when the complete CSI sequence is used. However, considerable high accuracy is reached when more than $75$\% of the input is employed. This accuracy of more than $95$\% should suffice for many practical purposes.

\section{Conclusion}
\label{sec:concl}
We described a system to predict a user's next cell. Formulating this prediction as a classification problem allowed us to apply \acp{SVM} on the last handover event and a CSI vector. This interesting combination of (i) long-term handover information and (ii) short-term CSI shows promising results for a synthetic Manhattan grid scenario and for a realistic radio map of Downtown Frankfurt. From these simulation results, we conclude that
\begin{enumerate*}
\item A realistic number of users is sufficient to train the system. In the studied scenarios, 100 users per path sufficed to build accurate classifiers.

\item \acp{SVM} predict the next cell substantially more accurately with CSI than with handover history alone. In the studied scenarios, our method more than doubled the accuracy of the classic handover history-based approach.

\item This high accuracy is already reached with a small fraction of the CSI input vector. In the studied scenarios, not more than 60\,\% of the CSI vector was required to reach 100\,\% prediction accuracy.
\end{enumerate*}
From these observations, we can conclude immediate practical benefits. Using only a small part of the input vector allows early but accurate prediction. Fast training makes the system reactive to changes in the environment. Finally, the prediction accuracy is high under various practical conditions. To this end, the proposed prediction framework is highly promising and deserves further study in realistic scenarios.

\section*{Acknowledgment}
We thank Lutz Ewe from Bell Labs, Stuttgart for supporting us with the Frankfurt radio map.
%\begin{acronym}
	\acrodef{SVM}{Support Vector Machine}
    \acrodef{PPT}{point-to-point}
    \acrodef{M2M}{Machine-to-Machine}
    \acrodef{BER}{Bit Error Rate}
    \acrodef{16-QAM}{16 Quadrature Amplitude Modulation}
    \acrodef{64-QAM}{64 Quadrature Amplitude Modulation}
    \acrodef{256-QAM}{256 Quadrature Amplitude Modulation}
    \acrodef{ACF}{Autocorrelation Function}
    \acrodef{ARQ}{Automatic Repeat Request}
    \acrodef{aDA}{advanced Dynamic Algorithm}
    \acrodef{ADC}{Analog-to-Digital Converter}
    \acrodef{APP}{Application layer}
    \acrodef{ASIC}{Application Specific Integrated Circuits}
    \acrodef{AWGN}{Additive White Gaussian Noise}
    \acrodef{BPSK}{Binary Phase Shift Keying}
    \acrodef{CBR}{Constant Bit Rate}
    \acrodef{CC}{Chase Combining}
    \acrodef{CDF}{Cumulative Distribution Function}
    \acrodef{CDMA}{Code-Division-Multiple-Access}
    \acrodef{CIC}{Cascaded Integrator-Comb}
    \acrodef{CIF}{Common Intermediate Format}
    \acrodef{CNR}{Channel Gain-to-Noise Ratio}
    \acrodef{CRC}{Cyclic Redundancy Check}
    \acrodef{CSI}{Channel State Information}
    \acrodef{CS}{Carrier Sensing}
    \acrodef{DAC}{Digital-to-Analog Converter}
    \acrodef{DCT}{Discrete Cosine Transform}
    \acrodef{DIV}{Distortion In Interval}
%!! don't use DLC acronym in this paper (it's defined in the intro, simply use "DLC" instead)
%    \acrodef{DLC}{Data Link Control layer}
    \acrodef{DSP}{Digital Signal Processor}
    \acrodef{EGC}{Equal Gain Combining}
    \acrodef{FDMA}{Frequency Division Multiple Access}
    \acrodef{FEC}{Forward Error Correction}
    \acrodef{FER}{Frame Error Rate}
    \acrodef{FS}{Frame Selection}
    \acrodef{FIFO}{First-In-First-Out}
    \acrodef{FPGA}{Field Programmable Gate Array}
    \acrodef{FSC}{Frame Check Sequences}
    \acrodef{GMSK}{Gaussian Minimum Shift Keying}
    \acrodef{GoP}{Group of Pictures}
    \acrodef{GSR}{GNU Software Radio}
    \acrodef{HTTP}{Hypertext Transfer Protocol}
    \acrodef{HTML}{Hypertext Mark-up Language}
    \acrodef{ICI}{Inter-carrier Interference}
    \acrodef{IEEE}{Institute of Electrical and Electronics Engineers, Inc.}
    \acrodef{IP}{Internet Protocol}
    \acrodef{IPC}{Inter-Process Communication}
    \acrodef{ISI}{Inter-symbol Interference}
    \acrodef{ISM}{industrial, scientific and medical}
    \acrodef{LLC}{Logical Link Control layer}
    \acrodef{LOS}{Line Of Sight}
    \acrodef{MAC}{Medium Access Control}
    \acrodef{MCM}{Multi Carrier Modulation}
    \acrodef{MIMO}{Multiple-Input Multiple-Output}
    \acrodef{MISO}{Multiple-Input Single-Output}
    \acrodef{MPEG}{Moving Pictures Expert Group}
    \acrodef{MOS}{Mean Opinion Score}
    \acrodef{MRC}{Maximum Ratio Combining}
    \acrodef{SC}{Selection Combining}
    \acrodef{MSB}{Most Significant Bit}
    \acrodef{MSS}{Maximum Segment Size}
    \acrodef{MTU}{Maximum Transmission Unit}
    \acrodef{NAV}{Network Allocation Vector}
    \acrodef{NCR}{Non-Cooperative Relaying}
    \acrodef{NLOS}{Non-Line Of Sight}
    \acrodef{NPS}{Network Path Selection}
    \acrodef{OFDM}{Orthogonal Frequency Division Multiplexing}
    \acrodef{OS}{Operating System}
    \acrodef{OR}{Opportunistic Relaying\slash{}Routing}
    \acrodef{PDF}{Probability Density Function}
    \acrodef{PDU}{Protocol Data Unit}
    \acrodef{PER}{Packet Error Rate}
    \acrodef{PHY}{Physical layer}
    \acrodef{PSC}{Packet Selection Combining}
    \acrodef{PSNR}{Peak Signal-to-Noise Ratio}
    \acrodef{QCIF}{Quarter CIF}
    \acrodef{QoS}{Quality of Service}
    \acrodef{QPSK}{Quadrature Phase Shift Keying}
    \acrodef{RCPC}{Rate-Compatible Punctured Convolutional}
    \acrodef{RF}{Radio Frequency}
    \acrodef{RMS}{root mean square}
    \acrodef{RTT}{Round Trip Time}
    \acrodef{SDF}{Selection Decode-and-Forward}
    \acrodef{SDR}{Software Defined Radio}
    \acrodef{SR}{Software Radio}
    \acrodef{SEP}{Symbol Error Probability}
    \acrodef{SCM}{Single Carrier Modulation}
    \acrodef{SDC}{Selection Diversity Combining}
    \acrodef{SIFS}{Short Inter-Frame Space}
    \acrodef{SNR}{Signal-to-Noise Ratio}
    \acrodef{TCP}{Transmission Control Protocol}
    \acrodef{TDMA}{Time Division Multiple Access}
    \acrodef{UDP}{User Datagram Protocol}
    \acrodef{USRP}{Universal Software Radio Peripheral}
    \acrodef{VBR}{Variable Bit Rate}
    \acrodef{VQM}{Video Queue Management}
    \acrodef{WLAN}{Wireless Local Area Network}
    \acrodef{WMAN}{Wireless Metropolitan Area Network}
    \acrodef{WSN}{Wireless Sensor Network}
    \acrodef{WT}{Wireless Terminal}
    \acrodef{WWW}{World-Wide-Web}
%\end{acronym}

\bibliographystyle{IEEEtran}
\bibliography{CSIFingerprinting}
%\bibliography{biblio-book-2011-03-13}
\end{document}